\documentclass[twocolumn,showpacs,preprintnumbers,amsmath,amssymb,aps]{revtex4-1}
\usepackage[dvips]{graphicx}
\usepackage{color}
\usepackage[all]{xy}
\usepackage{ulem}
 
\def\setR{\mathbb{R}} 
\def\setN{\mathbb{N}}

\def\nlb{\nolinebreak}

\newcommand{\norm}[1]{\parallel\!#1\!\parallel}

\newcommand{\formsp}[2]{\langle \, #1,  #2 \,\rangle}

\newcommand{\sss}[1]{\scriptscriptstyle #1}

\begin{document}

\title{Explicit vector spherical harmonics on the 3-sphere}

\author{J. Ben Achour$^1$}
\author{E. Huguet$^1$}
\author{J.Queva$^2$}
\author{J. Renaud$^3$}
\affiliation{$1$ - Universit\'e Paris Diderot-Paris 7, APC-Astroparticule et Cosmologie (UMR-CNRS 7164), 
Batiment Condorcet, 10 rue Alice Domon et L\'eonie Duquet, F-75205 Paris Cedex 13, France.\\
$2$ - Equipe Physique Th\'eorique, Projet COMPA, SPE (UMR 6134), 
Universit\'e de Corse, BP52, F-20250,  Corte, France.\\
$3$ - Universit\'e Paris-Est, APC-Astroparticule et Cosmologie (UMR-CNRS 7164), 
Batiment Condorcet, 10 rue Alice Domon et L\'eonie Duquet, F-75205 Paris Cedex 13, France.
} 
\email{benachou@apc.univ-paris7.fr,\\huguet@apc.univ-paris7.fr, \\queva@univ-corse.fr,\\jacques.renaud@apc.univ-paris7.fr}

\date{\today}


\begin{abstract}
We  build a family of explicit one-forms on $S^3$ which are shown to form a new complete set of eigenmodes for the 
Laplace-de Rahm operator. 
\end{abstract}

\maketitle

\section{Introduction}\label{SEC-Introduction}

The problem of the determination of tensorial modes for the Laplacian on spheres is important in many areas of Physics.
It has been tackled in the past as 
part of various works in different fields. As a consequence the references on 
that subject may not be so easy to find. Let us summarize the specific case of vector modes on $S^3$. It has been considered 
from different point of view: Adler \cite{Adler} uses an explicit embedding in a larger
space, Gerlach and Sengupta \cite{GerlachSengupta} solve the eigenvector equation in the hyperspherical coordinates system, and 
Jantzen \cite{Jantzen} makes use of the identification between $S^3$ and SU$(2)$ to obtain general properties using group theoretical 
methods. The main results of these works (and others) has been summarized and extended
by Rubin and Ord\'o\~{n}ez \cite{RubinOrdI}-\cite{RubinOrdII} and also Copeland and Toms \cite{CopelandToms}. However, 
the vector modes do not appear in closed form in all these works. On the contrary, the specific case of transverse vector 
modes are given explicitly by Higuchi \cite{Higuchi} in relation with the representations of the SO$(4)$ group.
In the present paper new vectors, or more exactly one-forms, modes for the Laplace-de Rahm operator on $S^3$ are built, in 
a compact manner, upon scalar modes using differential geometric methods.

In our work, the basic idea for finding the modes is reminiscent of that
used for the vector modes for the Laplacian on $S^2 \subset \setR^3$.
That is, starting from a scalar mode one
builds up its gradient ($\overrightarrow\nabla\Phi$), then the curl of the scalar field times a constant 
vector ($\overrightarrow\nabla\times\overrightarrow a \Phi$), then the curl of this vector 
($\overrightarrow\nabla\times\overrightarrow\nabla\times\overrightarrow a \Phi$). Here the gradient will be replaced by the exterior
derivative, the curl by the operator $*d$, and the constant vector $\overrightarrow a$ by a Killing vector of unit norm.

The structure of the paper is as follows. Notations, conventions and useful formulas are collected in Sec. \ref{SEC-Notations}. The 
scalar modes in the Hopf coordinates and some of their properties are reminded in Sec. \ref{SEC-ScalHarmTorCoord}. The main result 
is detailed in Sec. \ref{SEC-OneFormHarm}.  

\subsection*{Notations and conventions}\label{SEC-Notations}
Our notations follow those of \cite{Fecko}.
The co-vector associated with a one-form $\alpha$ is denoted $\widetilde\alpha\nlb:=\nlb\sharp\alpha$, and 
the one-form associated with a vector $\xi$ is also denoted $\widetilde\xi:=\flat\xi$.
The interior product with a vector $\xi$ is denoted by $i_\xi$ and the exterior product with a one-form $\alpha$
by $j_\alpha \beta:=\alpha \wedge\beta$, for any p-form $\beta$. We note  that the operators 
$i$ and $j$ are nilpotent. For a p-form 
$\alpha$ we define $\hat\eta$ by $\hat\eta\alpha = (-1)^p\alpha$. The 
scalar product for p-forms $\alpha$, $\beta$ on $S^3$ is 
$\formsp{\alpha}{\beta} := \int_{\sss S^3} \alpha \wedge \ast \beta.
$
The Laplace-de~Rahm operator is defined as $\Delta\nlb:=-(\delta d\nlb + \nlb d \delta)$. 

For convenience we reproduce here the relations we repeatedly used in calculations on $S^3$, they are:
$i\ast\nlb = \nlb\ast j \hat\eta$, $\ast\hat\eta\nlb =\nlb -\hat\eta\ast$, $\delta\ast\nlb =\nlb -\ast d \hat\eta$,  
$d\ast\nlb=\nlb\ast\delta\hat\eta$, $[\Delta, \ast] = 0$; for a Killing 
vector field $\xi$, one has $[\mathcal{L}_\xi, X] = 0$ for $X=\ast,d, \sim$;  for $\alpha, \beta$ p-forms and $\gamma$ a 
one-form one has $\formsp{j_\gamma\alpha}{\beta}\nlb=\nlb\formsp{\alpha}{i_{\widetilde\gamma}\beta}$.

\section{Scalar harmonics and  Hopf coordinates}\label{SEC-ScalHarmTorCoord}
Let us first introduce the  Hopf coordinates defined on the unit sphere $S^3$  by
\begin{equation}\label{EQ-TorroidalCoord}
 \begin{cases}
  x^1 =& \sin \alpha \cos \varphi\\
  x^2 =& \sin \alpha \sin \varphi\\
  x^3 =& \cos \alpha \cos \theta\\
  x^4 =& \cos \alpha \sin \theta 
 \end{cases}
\end{equation}
with $\alpha \in [0, \pi/2]$, $\theta, \varphi \in [0, 2\pi[$. In this system the metric element on $S^3$ reads 
$ds^2 =  d\alpha^2 + \cos^2\alpha\, d\theta^2 + \sin^2\alpha\, d\varphi^2$, in the coordinates basis or  
$ds^2 =  (e^\alpha)^2 + (e^\theta)^2 + (e^\varphi)^2$ in the orthonormal, direct co-frame $e^\alpha:= d\alpha$, 
$e^\theta:= \cos\alpha\, d\theta$, $e^\varphi := \sin\alpha \,d\varphi$.

We now consider the normalized scalar modes for the Laplace-de Rahm operator $\Delta := - (d\delta + \delta d)$ on $S^3$ 
(see \cite{LachiezeCaillerie} for instance). They 
satisfy the eigenvalues equation
\begin{equation}
\Delta \Phi_i = \lambda_i \Phi_i,
\end{equation}
where $\Phi_i$ stands for the modes corresponding to the eigenvalue $\lambda_i = -L(L+2)$, with $L \in \setN$, the 
index $i$ is a shorthand for the indexes needed to label the modes.
In the  system 
\eqref{EQ-TorroidalCoord}, for instance, the modes are labeled by the
three numbers  $(L, m^{~}_+, m^{~}_-)$,  where $m^{~}_+$, $m^{~}_-$  are such that
$\vert m_{\pm} \vert \leqslant \frac{L}{2}$, and
$\frac{L}{2} - m_\pm \in \setN$, they read
\begin{align}
    \Phi_i\label{EQ-TorroidalFunc}
    &= T_{{\sss L},m^{~}_+, m^{~}_-}(\alpha,\varphi,\theta) \\
    &:= C_{{\sss L},m^{~}_+,m^{~}_-} e^{i \left(S \varphi + D \theta\right)}
    (1-x)^{\frac{S}{2}} (1+x)^{\frac{D}{2}}
    P_{\frac{\sss L}{2}-m^{~}_+}^{(S, D)}(x),
    \nonumber
\end{align}
in which $P^{(a,b)}_n$ is a Jacobi polynomial, $x :=\cos 2\alpha$, $S\nlb:=\nlb m^{~}_+ + m^{~}_-$, $D\nlb:=\nlb m^{~}_+-m^{~}_-$ and
\begin{equation*}
C_{{\sss L},m^{~}_+,m^{~}_-} := \frac{1}{2^{m^{~}_+}\pi} \sqrt{\frac{L+1}{2}}
\sqrt{\frac{(L/2 + m^{~}_+)!(L/2 - m^{~}_+)!}{(L/2 + m^{~}_-)!(L/2 - m^{~}_-)!}}.
\end{equation*}
Let us finally introduce the two Killing vectors
\begin{align}
\xi : =& X_{12} + X_{34} = \partial_\varphi + \partial_\theta,\label{EQ-DefKillingXi}\\
\xi' : =& X_{12} - X_{34} = \partial_\varphi - \partial_\theta,\label{EQ-DefKillingSigma}
\end{align}
where $X_{ij} := x_i\partial_j - x_j\partial_i$ are the generators of the so$(4)$ algebra. Using the expression of the 
$S^3$-metric we see that $\norm{\xi}=\norm{\xi'}=1$. 
A straightforward calculation  shows that the associated one-forms to these Killing vectors are 
eigenvectors of the operator $\ast d$, one has
\begin{equation}
    \label{EQ-*dxi}
    \ast d \widetilde\xi = -2 \widetilde\xi,~~~%
    \ast d \widetilde\xi' = +2 \widetilde\xi'.
\end{equation}

In addition,  the scalar modes $\Phi_i$ are eigenmodes of $\xi$ and $\xi'$
(seen as differential operators), one has
\begin{equation*}
\xi(\Phi_i) = \mu_i \Phi_i,~~~\xi'(\Phi_i) = \nu_i \Phi_i,
\end{equation*}
where $\mu_i = + 2 i m^{~}_+,~~~\nu_i = - 2 i m^{~}_-$.

\section{One-form harmonics}\label{SEC-OneFormHarm}
In this section, we are interested in the eigenvalue equation $\Delta \alpha = \lambda \alpha$, where $\alpha$ is a one-form. 
The space of eigenvectors for this equation is the direct sum of two orthogonal subspaces containing respectively the exact and the 
co-exact one-forms. The exact one-forms are given by the exterior derivatives of the scalar modes, their eigenvalues are
$-L(L+2)$,  $L\in\setN \diagdown \{0\}$, the dimension of the associated proper subspace $\mathcal{E}^{\sss E}_{\sss L}$ is
$d^{\sss E}=(L+1)^2$ \cite{CopelandToms}. For the co-exact one-forms
the eigenvalues are known to be $-L^2, L\in\setN \diagdown \{0, 1\}$ and the dimension of the associated proper 
subspaces $\mathcal{E}^{\sss CE}_{\sss L}$ is $d^{\sss CE} = 2 (L-1)(L+1)$ \cite{CopelandToms}. Here, we will build up an explicit 
new orthonormal basis of co-exact eigenmodes. Our strategy will be to exhibit a family of modes associated to the eigenvalue $-L^2$, 
to show their orthogonality, and finally to check that their number is precisely the dimension of the proper
subspace  $\mathcal{E}_{\sss L}^{\sss CE}$.

\subsection{Definition of the modes}\label{SEC-DefModes}

Using the scalar modes let us define:
\begin{align*}
    A_i  &:= d\Phi_i, && \\
    B_i  &:= \ast d\Phi_i\widetilde{\xi}, &
    B'_i &:= \ast d\Phi_i\widetilde{\xi'}, \\
    C_i  &:= \ast d B_i, &
    C'_i &:= \ast d B'_i,
\end{align*}
in which $A_i$ is an exact one-form while $B_i$, $C_i$, $B'_i$
and $C'_i$ are co-exact one-forms.
Let us, in addition, consider the combinations
\begin{equation}\label{EQ-DefEiE'i}
    E_i := (L + 2) B_i + C_i,~~~ E'_i := (L + 2) B'_i - C'_i,
\end{equation}
where $i = (L, m^{~}_+, m^{~}_-)$  (respectively $(L, m'_+, m'_-)$).

\subsection{Statement of the results}

The following properties hold:
\begin{enumerate}
\item The one-forms $E^{~}_{{\sss L},m^{~}_+,m^{~}_-}$ and  $E'_{{\sss L},m'_+,m'_-}$   satisfy
\begin{align*}
 \Delta E^{~}_{{\sss L},m^{~}_+,m^{~}_-} &=-L^2E^{~}_{{\sss L},m^{~}_+,m^{~}_-},\\
 \Delta E'_{{\sss L},m'_+,m'_-} &=-L^2 E'_{{\sss L},m'_+,m'_-},
 \end{align*}
 for $L\geqslant2$.
\item The family of one-forms
\begin{equation*}
\begin{cases}
E^{~}_{{\sss L},m^{~}_+,m^{~}_-} , L\geqslant2, \vert m^{~}_+ \vert \leqslant\frac{L}{2}-1, \vert m^{~}_- \vert \leqslant\frac{L}{2},&\\
E'_{{\sss L},m'_+,m'_-} , L\geqslant2, \vert m'_- \vert \leqslant\frac{L}{2}-1, \vert m'_+ \vert \leqslant\frac{L}{2},
\end{cases}
\end{equation*}
once normalized, form 
an orthonormal basis of the corresponding proper subspace of co-exact one-forms $\mathcal{E}_{\sss L}^{\sss CE}$. 
\item The whole set of one-forms: exact $\{A_{{\sss L},m^{~}_+,m^{~}_-}\}$, for $L\geqslant\nolinebreak1$, and co-exact 
$\{E^{~}_{{\sss L},m^{~}_+,m^{~}_-}, E'_{{\sss L},m'_+,m'_-}\}$, as above, form a complete 
orthonormal set of modes for the Laplace-de Rahm operator on $S^3$.
\end{enumerate}

Finally, let us note that the modes $E^{~}_i$ and $E'_i$ can be recast under a vectorial form which is reminiscent
of the results for the two-sphere reminded in the introduction Sec. \ref{SEC-Introduction}, namely
\begin{align*}
\overrightarrow E_i &= (L + 2) \left(\overrightarrow\nabla\times(\Phi_i\overrightarrow\xi^{~})\right) + 
\overrightarrow\nabla\times\overrightarrow\nabla\times(\Phi_i\overrightarrow\xi),\\
\overrightarrow E'_i &= (L + 2) \left(\overrightarrow\nabla\times(\Phi_i\overrightarrow\xi')\right) - 
\overrightarrow\nabla\times\overrightarrow\nabla\times(\Phi_i\overrightarrow\xi').
\end{align*}

\subsection{Proof of the first property}\label{SEC-Proof1}
We first note the following property: 
    if a co-exact one-form $\alpha$ is an eigenmode of
    $\ast d$ with the eigenvalue $\sigma$ then
    $\alpha$ is an eigenmode of $\Delta$ with the
    eigenvalue $\sigma^2$. Indeed, one has:
\begin{equation*}
    \Delta \alpha 
    = - \delta d \alpha
    = -\ast d \ast \hat\eta  d \alpha
    = -\ast d \ast  d \alpha
    = - \sigma^2\alpha.
\end{equation*}
We consequently first consider the operator $\ast d$.

The action of $\ast d$ on $B_i$ is just the definition of $C_i$ (Sec. \ref{SEC-DefModes}). It remains to 
determine $\ast d C_i$. From the definition of $B_i$ and $C_i$ one has, using (\ref{EQ-*dxi})  
and $\Delta\Phi_i = -  \delta d \Phi_i$,
\begin{align}
C_i =& \ast dB_i\nonumber\\
=& \ast d \ast dj_{\widetilde\xi}\Phi_i\nonumber\\
=& -\ast d \ast j_{\widetilde\xi} d\Phi_i + \ast d\Phi_i*d\widetilde\xi \nonumber\\
=& *d i_\xi *d\Phi_i + \ast d\Phi_i (-2)\widetilde\xi\nonumber\\
=& \ast (\mathcal{L}_\xi - i_\xi d ) \ast d\Phi_i - 2 B_i\nonumber\\
=& d\mathcal{L}_\xi\Phi_i - \ast i_\xi d \ast d\Phi_i - 2 B_i\nonumber\\
=& d\mu_i\Phi_i + j_{\widetilde\xi} \delta d \Phi_i - 2 B_i\nonumber\\
C_i=& \mu_i A_i -\lambda_i\Phi_i\widetilde\xi - 2 B_i. \label{EQ-ExpDeCi}
\end{align}
Applying $\ast d$ to the above expression of $C_i$, taking into account the exactness of $A_i$, leads to
\begin{equation}\label{EQ-*dC}
\ast d C_i = -\lambda_i B_i - 2 C_i.
\end{equation}
This relation together with the definition of $C_i$, namely  $\ast dB_i =: C_i$, is a closed system of equations which can be 
diagonalized to obtain eigenmodes of $\ast d$.  A straightforward calculation with $\lambda_i$ replaced by its value $-L(L+2)$ 
leads to
\begin{align}
\ast d E_i =&  + L E_i,\label{EQ-*dEi}\\
\ast d F_i =& - (L+2) F_i.
\end{align}
where $E_i := (L + 2) B_i + C_i$ is the combination given in Sec \ref{SEC-DefModes} and  $F_i:=  L B_i - C_i$ is another 
set of modes that we do not need to consider further.
We then apply the property quoted at the beginning of this section to the co-exact one-form $E_i$, this leads to the result
\begin{equation}\label{EQ-DeltaEi}
\Delta E_i = - L^2 E_i. 
\end{equation}

A completely analogous calculation using $B'_i$ and $C'_i$ in place of $B_i$ and $C_i$ shows that the $E'_i$'s 
are eigenmodes of $\ast d$ with opposite eigenvalues and thus of $\Delta$ with the same eigenvalues. This completes 
the proof of the first property.

We thus have two families of modes with completely similar properties, the 
formulas for the $E'_i$'s being obtained through the same calculations, 
but using primed quantities ($B'_i$, $C'_i$, \ldots), as those leading to the 
results for the $E_i$'s . Consequently we will consider the 
family $\{E_i\}$ and only state the results for the family $\{E'_i\}$.

The formula (\ref{EQ-DeltaEi}) is valid for any $L\geqslant 0$,  nevertheless the  eigenvalue $L^2 = 0$ is 
excluded because it corresponds to harmonic one-forms whose set on a Riemannian 
manifold with positive Ricci curvature is known 
to only contains the null one-form  (Bochner's theorem). As we will see in the 
forthcoming paragraph this is in accordance with 
the norm of the corresponding eigenvector $E_{000}$ ($L=0$) which vanishes. 
Moreover, we will see that $E_i$ and $E'_i$ also vanish  for $L=1$. This 
explains the additional condition $L \geqslant 2$ for the eigenvalues of 
the modes $E_i$ and $E'_i$.

\subsection{Scalar products and norms}\label{SEC-ScalarProductsNorms}
From their definition \eqref{EQ-DefEiE'i} the scalar products between the
$E_i$'s and $E'_i$'s can be deduced from those between the $B_i$'s, $C_i$'s,
$B'_i$'s and $C'_i$'s, which we compute hereafter. 

We begin by the scalar product between the $B_i$'s which reads
\begin{align*}
\formsp{B_i}{B_j} =& \formsp{\ast d j_{\widetilde\xi}\Phi_i}{\ast d j_{\widetilde\xi}\Phi_j}\\
 =& \formsp{\Phi_i}{i_\xi \delta d j_{\widetilde\xi}\Phi_j}.
\end{align*}
In order to calculate the term in the bracket we observe that $\delta d j_{\widetilde\xi}\Phi_j= C_j$,
using the expression (\ref{EQ-ExpDeCi}) we now calculate $i_\xi C_j$, one has
\begin{align*}
i_\xi C_j =& i_\xi(\mu_j d\Phi_j -\lambda_j\Phi_j\widetilde\xi - 2 \ast d j_{\widetilde\xi}\Phi_j)\\
=&\mu_j (\mathcal{L}_\xi - d i_\xi ) \Phi_j -\lambda_j\Phi_j\norm{\xi}^2 \\
- &2 (i_\xi \Phi_j \ast d \widetilde\xi - i_\xi\ast j_{\widetilde\xi}d\Phi_j)\\
=&\mu_j \xi(\Phi_j) - \lambda_j\Phi_j\norm{\xi}^2 + 4\Phi_j\norm{\xi}^2\\
=&(\mu_j^2 - \lambda_j + 4)\Phi_j,
\end{align*}
where we used $i_\xi\ast j_{\widetilde\xi}= 0$ and $\norm{\xi}=1$. Finally,
\begin{equation}\label{EQ-psBiBj}
\formsp{B_i}{B_j} =(\mu_i^2 - \lambda_i + 4)\delta_{ij},
\end{equation}
where $\delta_{ij}$ has to be interpreted as the product of the Kronecker symbols of the various numbers 
labeling the modes.
For $i= (L, m^{~}_+, m^{~}_-)$ and $j =  (K, n_+, n_-)$ one has
$\delta_{ij} = \delta_{\sss LK}\delta_{m^{~}_+n_+}\delta_{m^{~}_-n_-}$.

The product $\formsp{B_i}{C_j}$ reads
\begin{align*}
\formsp{B_i}{C_j}=&\formsp{\ast d j_{\widetilde\xi}\Phi_i}{C_j}\\
=&\formsp{\Phi_i}{i_\xi \ast d C_j}.
\end{align*}
Keeping in mind the previous calculations, the r.h.s. of the bracket reads
\begin{align*}
i_\xi \ast d C_j =& i_\xi \ast d (\mu_j A_j -\lambda_j\Phi_j\widetilde\xi - 2 B_j)\\
=& -\lambda_j i_\xi \ast d\Phi_j\widetilde\xi - 2 i_\xi C_j\\
=& -\lambda_j i_\xi(-\ast j_{\widetilde\xi} d\Phi_j + \Phi_j \ast d\xi) \\ 
-&2 i_\xi (\mu_j A_j -\lambda_j\Phi_j\widetilde\xi - 2 \ast d j_{\widetilde\xi}\Phi_j)\\
=& -\lambda_j i_\xi( i_\xi \ast  d\Phi_j + \Phi_j (-2 \widetilde\xi) ) \\ 
-&2 (\mu_j^2 \Phi_j-\lambda_j\norm{\xi}^2 \Phi_j -2 i_\xi\ast d j_{\widetilde\xi}\Phi_j)\\
=&-2 (\mu_j^2 - 2 \lambda_j\norm{\xi}^2  + 4 \norm{\xi}^2)\Phi_j.
\end{align*}
Finally, with $\norm{\xi} = 1$ we obtain
\begin{equation}\label{EQ-psBiCj}
\formsp{B_i}{C_j} = -2(\mu_i^2 - 2\lambda_i + 4)\delta_{ij}.
\end{equation}

The product between the $C_i$'s can be calculated using the results (\ref{EQ-psBiBj}) and (\ref{EQ-psBiCj}), one has
\begin{align*}
\formsp{C_i}{C_j}=&\formsp{\ast d B_i}{\ast d B_j}\\
=&\formsp{B_i}{\delta d B_j}\\
=&-\formsp{B_i}{\Delta B_j}\\
=&-\lambda_j\formsp{B_i}{B_j} - 2 \formsp{B_i}{C_j},
\end{align*}
from which we obtain
\begin{equation}
    \label{EQ-psCiCj}
    \formsp{C_i}{C_j}
    = \left[(4-\lambda_i)\mu_i^2 
        + (\lambda_i - 12)\lambda_i
        + 16\right]\delta_{ij}.
\end{equation}

From the scalar products (\ref{EQ-psBiBj}-\ref{EQ-psCiCj}) and the definition of
the modes \eqref{EQ-DefEiE'i}
a straightforward calculation leads to

\begin{equation*}
    \formsp{E_i}{E_j} = \norm{E_i}^2 \delta_{ij}, 
\end{equation*}
with here $i = (L, m^{~}_+, m^{~}_-)$ and in which the squared norm is given by:
\begin{equation}\label{EQ-NormE_i}
    \norm{E_i}^2 = 2L(L+1) (L^2 - 4 m^2_+), 
\end{equation}
where we have used the values of $\lambda_i = -L(L+2)$ and $\mu_i =
2\, i \, m^{~}_+$.

The norm of the eigenvector 
$E_{{\sss L},m^{~}_+,m^{~}_-}$ (\ref{EQ-NormE_i}),  vanishes for $m^{~}_+ = \pm\frac{L}{2}$, these values, which correspond 
to the boundaries of the spectrum, 
are thus excluded. For the value $L=1$, the two values allowed for $m^{~}_+$ are on the boundaries, 
the value $L=1$ 
is thus excluded.

Finally, a completely analogous calculation using the primed quantities ($B'_i$, $C'_i$,\ldots) leads to the same results as above for 
the modes $E'_i$ in which $\mu_i$ is replaced by $\nu_i$ that is $m^{~}_+$ is
replaced by $m^{~}_-$ in the squared norm of $E'_i$. 

\subsection{Proof of the second property}\label{SEC-Proof2}
The dimension of the proper subspace associated to a
given eigenvalue is given by its degeneracy. Following \cite{CopelandToms} (appendix B) the degeneracy of the eigenvalue 
$-L^2$ is 
$d^{\sss CE}\nlb=\nlb2(L-1)(L+1)$.  
Now, the results of Sec. \ref{SEC-ScalarProductsNorms}
show that for a given eigenvalue the number of eigenvectors $E_i$ is given
by the number of values for $m^{~}_+$ which correspond to a non-null eigenvector times the number of 
possible values for $m^{~}_-$. Then, for the  eigenvalue $-L^2$ the number of $E^{~}_{{\sss L},m^{~}_+,m^{~}_-}$ 
eigenvectors is $(L-1)(L+1)$. This is also the number of eigenvectors $E'_{{\sss L},m'_+,m'_-}$ for the 
same eigenvalue. Consequently, to prove the second property it is sufficient to show that the two families $E_i$ and $E'_i$
are orthogonal. This is a consequence of the fact that   the $\ast d$ operator is symmetric, namely 
\begin{align*}
\formsp{\ast dE_i}{E'_j}=&\formsp{E_i}{\delta\ast E'_j}\\
=&\formsp{E_i}{\ast d \ast \hat\eta \ast E'_j}\\
=&\formsp{E_i}{\ast d \ast \ast E'_j}\\
=&\formsp{E_i}{\ast d E'_j}.
\end{align*}
It follows that two eigenmodes corresponding to two different eigenvalues of $\ast d$ are orthogonal. Now, as quoted in 
the Sec. \ref{SEC-Proof1} the eigenvalues corresponding to
$E_i$ and $E'_i$ are respectively positive and negative integers and consequently have no value in common. Thus
the modes $E_i$ and $E'_j$ are orthogonal. This completes the proof of the second property.

\subsection{Proof of the third property}\label{SEC-Proof34}
The proof of the third point is as follows. First note that the $A_i$'s are eigenmodes for the Laplace-de Rahm operator
\begin{align*}
\Delta A_i =& -(d\delta + \delta d) d\Phi_i\\
=&-d (\delta d + d\delta) \Phi_i\\
=& \lambda_i A_i.
\end{align*}
Then, we determine the scalar product of two $A_i$'s 
\begin{align*}
\formsp{A_i}{A_j}=& \formsp{\Phi_i}{\delta d\Phi_j}\\
=& \formsp{\Phi_i}{-\lambda_j \Phi_j}\\
=& -\lambda_i \delta_{ij}.
\end{align*}

The family $A_i$ is thus a complete set of exact eigenmodes on $S^3$ since the 
dimension of each proper subspace is $(L+1)^2$. 

Finally, the proof of the last property amounts to show that the two sets of modes $\{A_{{\sss L},m^{~}_+,m^{~}_-}\}$ 
and $\{E^{~}_{{\sss L},m^{~}_+,m^{~}_-}, E'_{{\sss L},m'_+,m'_-}\}$ are orthogonal. This comes
from the fact that the members of the second family are co-exact, in fact for a co-exact form $\alpha$ one has:
\begin{equation*}
\formsp{A_i}{\alpha} = \formsp{\Phi_i}{\delta \alpha} = 0.
\end{equation*}

\section*{Acknowledgements}
The authors wish to thanks M.~Lachi\`eze-Rey for valuable discussions, and for the 
Killing vectors of unit norms defined in Sec. \ref{SEC-ScalHarmTorCoord}. 
Thanks must be addressed to A.~Higuchi for pointing us toward reference \cite{Higuchi}.

\end{document}